\documentclass[12pt]{iopart}
\usepackage{setstack}
\usepackage{amssymb}
\usepackage{graphicx}     % good graphics
\begin{document}
%---------------------------------------------------------------------------
\title[Vortex analogue for the equatorial geometry of the Kerr black hole]
{Vortex analogue for the equatorial geometry of the \\ Kerr black hole}
%---------------------------------------------------------------------------
\author{Matt Visser}
\address{
School of Mathematics, Statistics,  and Computer Science,\\
Victoria University of Wellington,\\ 
PO Box 600, Wellington, \\
New Zealand}
\ead{matt.visser@mcs.vuw.ac.nz}
%------------------------------------------------

%------------------------------------------------
\author{Silke Weinfurtner}
\address{
School of Mathematics, Statistics,  and Computer Science,\\
Victoria University of Wellington,\\ 
PO Box 600, Wellington, \\
New Zealand}
\ead{silke.weinfurtner@mcs.vuw.ac.nz}
%------------------------------------------------

%---------------------------------------------------------------------------
\date{23 December 2004; Revised 17 February 2005; \LaTeX-ed \today}
%---------------------------------------------------------------------------
%---------------------------------------------------------------------------
%---------------------------------------------------------------------------
%---------------------------------------------------------------------------
%---------------------------------------------------------------------------
\begin{abstract}
  The spacetime geometry on the equatorial slice through a Kerr black
  hole is formally equivalent to the geometry felt by phonons
  entrained in a rotating fluid vortex. We analyse this example of
  ``analogue gravity'' in some detail: First, we find the most general
  acoustic geometry compatible with the fluid dynamic equations in a
  collapsing/expanding perfect-fluid line vortex. Second, we
  demonstrate that there is a suitable choice of coordinates on the
  equatorial slice through a Kerr black hole that puts it into this
  vortex form; though it is not possible to put the entire Kerr
  spacetime into perfect-fluid ``acoustic'' form.  Finally, we discuss
  the implications of this formal equivalence; both with respect to
  gaining insight into the Kerr spacetime and with respect to possible
  vortex-inspired experiments, and indicate ways in which more general
  analogue spacetimes might be constructed.

\end{abstract}
%----------------------------------------------------------------------------

\pacs{02.40.Ma. 04.20.Cv, 04.20.Gz, 04.70.-s}

%---------------------------------------------------------------------------
\maketitle
%---------------------------------------------------------------------------

%--------------------------------------------------
\section{Introduction}
%---------------------------------------------------
\def\tr{\hbox{\rm tr}}
\def\implies{\Rightarrow}
\def\conv{\hbox{\rm conv}}
\def\Re{ {\cal R} }
\def\half{{{1\over2}}}
\def\d{{\mathrm{d}}}
%---------------------------------------------------

%-----------------------------------------------------------------------------
\def\lint{\hbox{\Large $\displaystyle\int$}} %needs \usepackage{amssymb}
\def\hint{\hbox{\Huge $\displaystyle\int$}}  %needs \usepackage{amssymb}
%-----------------------------------------------------------------------------

It is by now well-known that the propagation of sound in a moving
fluid can be described in terms of an effective spacetime geometry.
(See, for instance,~\cite{Unruh1,Unruh2,Visser1,Visser2,ABH,causal},
and references therein).
\begin{itemize}
\item 
In the geometrical acoustics approximation, this emerges in a
straightforward manner by considering the way the ``sound cones'' are
dragged along by the fluid flow, thereby obtaining the conformal class
of metrics (see, for instance,~\cite{ABH}):
\begin{equation}
g_{\mu\nu} \propto
\left[\begin{array}{c|c}{-\{c^2-h^{mn}\;v_n\;v_n}\} & -v_j\\
\hline
-v_i & h_{ij} 
\end{array}\right].
\end{equation}
Here $c$ is the velocity of sound, $v$ is the velocity of the fluid,
and $h_{ij}$ is the 3-metric of the ordinary Euclidean space of
Newtonian mechanics (possibly in curvilinear coordinates).

\item 
In the physical acoustics approximation we can go somewhat further: A
wave-equation for sound can be derived by linearizing and combining
the Euler equation, the continuity equation, and a barotropic equation
of state~\cite{Unruh1,Unruh2,Visser1,Visser2}.  For irrotational flow
this process leads to the curved-space d'Alembertian equation and in
particular now fixes the overall conformal factor. The resulting
acoustic metric in (3+1) dimensions is (see, for
instance,~\cite{Visser2,ABH}):
\begin{equation}
g_{\mu\nu} = \left({\rho\over c}\right)\;
\left[\begin{array}{c|c}{-\{c^2-h^{mn}\;v_n\;v_n}\} & -v_j\\
\hline
-v_i & h_{ij} 
\end{array}\right].
\end{equation}
Here $\rho$ is the density of the fluid. Sound is then described by a
massless minimally coupled scalar field propagating in this acoustic
geometry.  In the presence of vorticity a more complicated wave
equation may still be derived ~\cite{vorticity}. For frequencies large
compared to the local vorticity $\omega=||\nabla\times \vec v||$ that
more complicated wave equation reduces to the d'Alembertian and the
acoustic geometry can be identified in the usual manner. For more
details see~\cite{vorticity}.
\end{itemize}
Now because the ordinary Euclidean space ($h_{ij}$) appearing in these
perfect fluid acoustic geometries is Riemann flat, the 3-dimensional
space (given by the constant-time slices) in any acoustic geometry is
forced to be conformally flat, with 3-metric $g_{ij} = (\rho/c)\;
h_{ij}$. This constraint places very strong restrictions on the class
of (3+1)-dimensional geometries that can be cast into perfect fluid
acoustic form.  While many of the spacetime geometries of interest in
general relativity can be put into this acoustic form, there are also
important geometries that \emph{cannot} be put into this form; at
least for a literal interpretation in terms of flowing perfect fluid
liquids and physical sound.

In particular, the Schwarzschild geometry can, at the level of
geometrical acoustics, certainly be cast into this perfect fluid
acoustic form~\cite{Visser2}.  However, at the level of physical
acoustics (and working within the context of the
Painl\'eve--Gullstrand coordinates) there is a technical difficulty in
that the Euler and continuity equations applied to the background
fluid flow yield a nontrivial and unwanted (3+1) conformal factor
$\rho/c$~\cite{Visser2}. This overall conformal factor is annoying,
but it does not affect the sound cones, and so does not affect the
``causal structure'' of the resulting analogue
spacetime~\cite{causal}.  Furthermore any overall conformal factor
will not affect the surface gravity~\cite{Jacobson-Kang}, so it will
not directly affect the acoustic analogue of Hawking
radiation~\cite{Unruh1,Visser2}. So for most of the interesting
questions one might ask, the Schwarzschild geometry can for all
practical purposes be cast into acoustic form.

Alternatively, as we shall demonstrate below, the Schwarzschild
geometry can also be put into acoustic form in a different manner by
using isotropic coordinates --- see the appendix of this article for
details.  More generally, any spherically symmetric geometry (static
or otherwise) can be put into acoustic form --- details are again
provided in the appendix.  (This of course implies that the
Reissner--Nordstr\"om geometry can be put into acoustic form.)  In
particular, the FRW cosmologies can also be put into acoustic form,
both at the level of physical acoustics and at the level of
geometrical acoustics. In fact there are two rather different routes
for doing so: Either by causing the fluid to explode or by adjusting
the speed of
sound~\cite{FRW1,FRW2,Uwe1,Uwe2,Uwe3,Lidsey,Silke1,Silke2}.

There is however a fundamental restriction preventing the Kerr
geometry [and Kerr--Newman geometry] being cast into perfect fluid
acoustic form. It has recently been
established~\cite{Price,Valiente-Kroon1,Valiente-Kroon2} that no
possible time-slicing of the full Kerr geometry can ever lead to
conformally flat spatial 3-slices.  Faced with this fact, we ask a
more modest question: Can we at least cast a \emph{subspace} of the
Kerr geometry into perfect fluid acoustic form? Specifically, since we
know that the effective geometry of a generalized line vortex (a
``draining bathtub'' geometry) contains both horizons and ergosurfaces
\cite{Visser2,ABH}, one is prompted to ask: If we look at the
equatorial slice of the Kerr spacetime can we at least put that into
acoustic form? If so, then this opens the possibility of finding a
physically reasonable analogue model based on a vortex geometry that
might mimic this important aspect of the Kerr geometry.  Thus we have
three independent physics questions to answer:
\begin{itemize}
\item 
What is the most general perfect fluid acoustic metric that can [even
  in principle] be constructed for the most general [translation
  invariant] line vortex geometry?

\item 
Can the equatorial slice of Kerr then be put into this form? 
\\ 
(And if not, how close can one get?)

\item
By generalizing the analogue model to something more complicated than
a perfect fluid, can we do any better?
\end{itemize}
We shall now explore these three issues in some detail.

%--------------------------------------------------------------
\section{Vortex flow}
%--------------------------------------------------------------

%--------------------------------------------------------------
\subsection{General framework}
%--------------------------------------------------------------

The background fluid flow [on which the sound waves are imposed] is
governed by three key equations: The continuity equation, the Euler
equation, and a barotropic equation of state:
\begin{equation} \label{cy_eq}
        \frac{\partial \rho}{\partial t}+\nabla\cdot (\rho \; \vec{v})=0.
\end{equation}
\begin{equation} \label{eu_eq}
        \rho \left[  \frac{\partial \vec{v}}{\partial t}
        + (\vec{v} \cdot \nabla) \vec{v}  \right]=-\nabla p + \vec{f}.
\end{equation}
\begin{equation}
        p=p(\rho).
\end{equation}
Here we have included for generality an arbitrary external force $\vec
f$, possibly magneto-hydrodynamic in origin, that we can in principle
think of imposing on the fluid flow to shape it in some desired
fashion. From an engineering perspective the Euler equation is best
rearranged as
\begin{equation}
\label{euler-rearranged}
\vec{f} =  
\rho \left[
\frac{\partial \vec{v}}{\partial t} + (\vec{v} \cdot \nabla) \vec{v} \right]
+\nabla p,
\end{equation}
with the physical interpretation being that $\vec f$ is now telling
you what external force you would need in order to set up a specified
fluid flow.

%--------------------------------------------------------------
\subsection{Zero radial flow}
%--------------------------------------------------------------

Assuming now a cylindrically symmetric time-independent fluid flow
without any sinks or sources we have a line vortex aligned along the
$z$ axis with fluid velocity $\vec{v}$:
\begin{equation} 
\label{2d_v}
     \vec v(r) =   v_{\hat\theta}(r)\;\hat{\theta}.
\end{equation}
The continuity equation (\ref{cy_eq}) for this geometry is trivially
satisfied and calculating the fluid acceleration leads to
\begin{equation}
%\fl
\vec a = (\vec{v}\cdot \nabla) \vec{v} =
- \frac{v_{\hat \theta}(r)^2}{r} \; \hat{r}.
\end{equation}
Substituting this into the rearranged Euler equation
(\ref{euler-rearranged}) gives
\begin{equation}
\vec{f} = f_{\hat{r}} \; \hat{r} = 
\left\{ -\rho(r) \; \frac{v_{\hat{\theta}}(r)^2}{r} 
+c^2(r) \; \partial_{r} \rho(r)  \right\} \hat{r}
\end{equation} 
with the physical interpretation that:
\begin{itemize}

\item 
The external force $\vec{f}_{\hat{r}}$ must be chosen to precisely
cancel against the combined effects of centripetal acceleration and
pressure gradient.  The angular-flow is not completely controlled by
this external force, but is instead an independently specifiable
quantity. (There is only \emph{one} relationship between $f_r(r)$,
$\rho(r)$, $c(r)$, and $v_\theta(r)$, which leaves \emph{three} of
these quantities as arbitrarily specifiable functions.)
  
\item 
We are now considering the equation of state to be an \emph{output}
from the problem, rather than an \emph{input} to the problem. If for
instance $\rho(r)$ and $c^2(r)$ are specified then the pressure can be
evaluated from
 \begin{equation}
p(r) = \int c^2(r) \; {\d \rho\over\d r} \; \d r,
\end{equation}
and then by eliminating $r$ between $p(r)$ and $\rho(r)$, the EOS
$p(\rho)$ can in principle be determined.

\item 
For zero external force (and no radial flow), which
arguably is the most natural system to set up in the laboratory, we
have
\begin{equation}
{v_{\theta}^2(r)\over c^2(r)} =  r \; \partial_r \ln\rho(r),
\end{equation}
which still has two arbitrarily specifiable functions.

\item 
In the geometric acoustics regime the acoustic line-element for this
zero-source/ zero-sink line vortex is
\begin{eqnarray}
\d s^2 &\propto&
        -\left( c^2-v_{\hat\theta}^2\right) \d t^2        
        - 2 v_{\hat\theta} \; r \; \d\theta \; \d t
        +\d r^2 + r^2 \d\theta^2 +\d z^2.
\end{eqnarray}
In the physical acoustics regime the acoustic line-element is
\begin{equation}
%\fl
\d s^2=\left( \frac{\rho}{c} \right) 
        \Big[ -\left( c^2-v_{\hat\theta}^2\right) \d t^2  
        - 2 v_{\hat\theta} \; r \d\theta \;\d t
        +\d r^2 + r^2 \d\theta^2 +\d z^2
        \Big].
\end{equation}
The vortex quite naturally has a ergosurface where the speed of the
fluid flow equals the speed of sound in the fluid.  This vortex
geometry may or may not have a horizon --- since $v_{\hat r}$ is
identically zero the occurrence or otherwise of a horizon depends on
whether or not the speed of sound $c(r)$ exhibits a
zero. \footnote{The vanishing of $c(r)$ at a horizon is exactly what
  happens for Schwarzschild black holes (or their analogues) in either
  Schwarzschild or isotropic coordinate systems. This is rather
  different from the behaviour in Painleve--Gullstrand coordinates,
  but is a quite standard signal for the presence of a horizon.}  We
shall later see that this class of acoustic geometries is the most
natural for building analogue models of the equatorial slice of the
Kerr geometry.
\end{itemize}

%-----------------------------------------------------------------------
\begin{figure}[htb]
\label{F:vortex-non-collapse}
\vbox{
\vskip 20 pt

\centerline{{\includegraphics[width=3in]{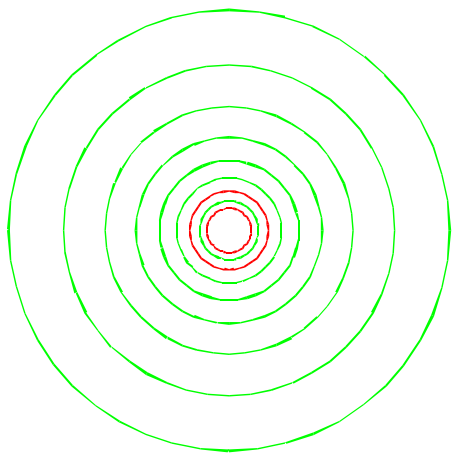}}}

\caption[Vortex geometry 1]{ {\sl A simple non-collapsing vortex
    geometry: The green circles denote the streamlines of the fluid
    flow. The outer red circle represents the ergosurface, where the
    fluid velocity reaches the speed of sound. The inner red circle
    (which need not exist in general) represents a possible [outer] 
    horizon where the speed of sound goes to zero.  \smallskip}} }
\end{figure}
%-----------------------------------------------------------------------

%--------------------------------------------------------------
\subsection{General analysis with radial flow}
%--------------------------------------------------------------

For completeness we now consider the situation where the vortex
contains a sink or source at the origin.  (A concrete example might be
the ``draining bathtub'' geometry where fluid is systematically
extracted from a drain located at the centre.)  Assuming now a
cylindrically symmetric time-independent fluid flow with a line vortex
aligned along the $z$ axis, the fluid velocity $\vec{v}$ is
\begin{equation} 
\label{2d_vb}
        \vec{v}=v_{\hat r}(r)\;\hat{r}+v_{\hat\theta}(r)\;\hat{\theta}.
\end{equation}
Wherever the radial velocity $v_{\hat r}(r)$ is nonzero the entire
vortex should be thought of as collapsing or expanding.  

%-----------------------------------------------------------------------
\begin{figure}[htb]
\label{F:vortex}
\vbox{
\vskip 20 pt

\centerline{{\includegraphics[width=3in]{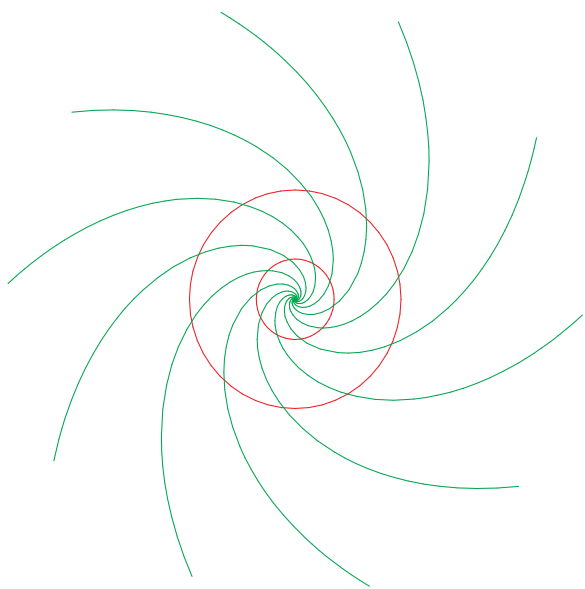}}}

\caption[Vortex geometry 2]{ {\sl A collapsing vortex geometry
    (draining bathtub): The green spirals denote streamlines of the
    fluid flow. The outer circle represents the ergosurface while the
    inner circle represents the [outer] event horizon.  \smallskip}} }
\end{figure}
%-----------------------------------------------------------------------

The continuity equation (\ref{cy_eq}) for this cylindrically symmetric
problem is
\begin{equation} \label{eu_ti_eq}
        \nabla \cdot (\rho \;\vec{v})=0,
\end{equation}
and the rearranged Euler equation (\ref{euler-rearranged}) for a
pressure $p$ which depends only on the radial-coordinate $r$ is
\begin{equation} \label{vec_f}
\vec f
=
\rho (\vec{v}\cdot \nabla) \vec{v} + c^2 \; {\partial}_r\rho \; \hat{r}.
\end{equation}
At this stage we note that $\rho$ is in general not an independent
variable.  Because equation (\ref{eu_ti_eq}) corresponds to a
divergence-free field, integration over any closed circle in the
two-dimensional plane yields
\begin{equation}
 \oint \rho(r)\;\vec{v}(r) \cdot \hat r \; \d s
= 2\pi\;\rho(r)\; v_{\hat r}(r)\; r = 2 \pi\; k_1.
\end{equation}
Then \emph{provided} $v_{\hat r}\neq 0$,
\begin{equation} \label{alpha_eq2}
         {\rho(r)} =\frac{k_1}{ r \;v_{\hat r}(r) }.
\end{equation}
Substituting into the rearranged Euler equation gives
\begin{equation}  
\vec f =
{k_1\over r v_{\hat r}} \; (\vec{v}\cdot \nabla) \vec{v}
+ c^2 \; {\partial}_r \left( {k_1\over r v_{\hat r}}\right) \; \hat{r},
\end{equation}
where $c$ is a function of $\rho$, and thus a function of $v_{\hat
  r}$.  This now completely specifies the force profile in terms of
the desired velocity profile, $ v_{\hat r}$, $ v_{\hat \theta}$, the
equation of state, and a single integration constant $k_1$.
Calculating the fluid acceleration leads to
\begin{equation}
\fl
\label{vec_a}
(\vec{v}\cdot \nabla) \vec{v} =
\left\{v_{\hat r}(r) \partial_r v_{\hat r}(r)
- \frac{v_{\hat \theta}(r)^2}{r} \right\}\hat{r}              
 + \left\{ v_{\hat r}(r) \partial_r v_{\hat \theta}(r)  +
 \frac{ v_{\hat r}(r)\;v_{\hat \theta}(r)}{r}  \right\} \hat{\theta},
\end{equation}
which can be rearranged to yield
\begin{equation}
(\vec{v}\cdot \nabla) \vec{v} =
\left\{ {1\over 2} \partial_r [v_{\hat r}(r)^2]
- \frac{v_{\hat \theta}(r)^2}{r} \right\}\hat{r}              
 + \left\{ {v_{\hat r}(r)\over r}  \partial_r [r\; v_{\hat \theta}(r)]  
  \right\} \hat{\theta}.
\end{equation}

Finally, decomposing the external force into radial and tangential
[torque-producing directions] we have
\begin{equation} \label{f_r}
f_{\hat r} =  \vec f \cdot \hat r = k_1 
\left\{ {1\over r v_{\hat r}} \;
\left[  {1\over 2} \partial_r [v_{\hat r}(r)^2]
- \frac{v_{\hat \theta}(r)^2}{r} \right]
+ c^2 \; {\partial}_r \left( {1\over r v_{\hat r}}\right)  \right\},
\end{equation}
and
\begin{equation} \label{f_theta}
f_{\hat\theta} =  \vec f \cdot \hat \theta  = k_1 
\left\{ {1\over r^2}  \partial_r [r\; v_{\hat \theta}(r)]  \right\}.
\end{equation}
The radial equation can undergo one further
simplification to yield
\begin{equation} \label{f_r_2}
f_{\hat r} = k_1 
\left\{ {1\over r}  \,
\left[ 1 - {c_s^2\over v_{\hat r}^2} \right] \,\partial_r v_{\hat r}
-  {c_s^2 + v_{\hat\theta}^2 \over r^2 \; v_{\hat r}} \right\}.
\end{equation}

\noindent Summarizing:
\begin{itemize}

\item 
In the geometric acoustics regime the acoustic line-element for the
most general [time-independent cylindrically symmetric collapsing/
  expanding] line vortex is
\begin{eqnarray}
\d s^2 &\propto&
        -\left( c^2-v_{\hat r}^2-v_{\hat\theta}^2\right) \d t^2  
        - 2v_{\hat r}\; \d r \; \d t
        - 2 v_{\hat\theta} \; r \; \d\theta \; \d t
\nonumber\\
&&\qquad
        +\d r^2 + r^2 \d\theta^2 +\d z^2.
\end{eqnarray}
We again reiterate that --- given a barotropic equation of state
$p(\rho)$ --- once the velocity profile $ v_{\hat r}$ and $ v_{\hat
  \theta}$ is specified, then up to a single integration constant
$k_1$, the density and speed of sound are no longer free but are fixed
by the continuity equation and the equation of state respectively.
Furthermore, the Euler equation then tells you exactly how much
external force is required to set up the fluid flow. There are still
\emph{two} freely specifiable functions which we can take to be the
two components of velocity.
  
\item 
In the physical acoustics regime the acoustic line-element is
\begin{eqnarray}
\d s^2&=&\left( \frac{\rho}{c} \right) 
        \Big[ -\left( c^2-v_{\hat r}^2-v_{\hat\theta}^2\right) \d t^2  
        - 2v_{\hat r}\; \d r \;\d t
        - 2 v_{\hat\theta} \; r \d\theta \;\d t
\nonumber\\
&&
\qquad\qquad
        +\d r^2 + r^2 \d\theta^2 +\d z^2
        \Big].
\end{eqnarray}
The major difference for physical acoustics is that for technical
reasons the massless curved-space Klein--Gordon equation
[d'Alembertian wave equation] can only be derived if the flow has zero
vorticity. This requires $v_{\hat\theta}(r) =k_2/r$ and hence
$f_{\hat\theta}=0$, so that the flow would be un-torqued.  More
precisely, the d'Alembertian wave equation is a good approximation as
long as the frequency of the wave is high compared to the vorticity.
However, in the presence of significant torque and vorticity, a more
complicated wave equation holds~\cite{vorticity}, but that wave
equation requires additional geometrical structure beyond the
effective metric, and so is not suitable for developing general
relativity analogue models.  (Though this more complicated set of
coupled PDEs is of direct physical interest in its own right.)~\footnote{We shall subsequently see that the ``equivalent Kerr vortex'' is not irrotational --- but the vorticity is proportional to the angular momentum, so there is a large parameter regime in which the effect of vorticity is negligible.}
%%% ***
\end{itemize}
There are several special cases of particular interest:
\begin{itemize}
\item[-] No vorticity, $\nabla\times\vec v=0$;
\item[-] No angular torque, $f_{\hat\theta}=0$;
\item[-] No radial flow $v_{\hat r}=0$;
\item[-] No radial force, $f_{\hat r}=0$;
\item[-] No external force, $\vec f = 0$;
\end{itemize}
which we will now explore in more detail.
%

%--------------------------------------------------------------
\paragraph{Zero vorticity/ zero torque:}
%--------------------------------------------------------------
If we assume zero vorticity, $\nabla\times\vec v=0$, the above
calculation simplifies considerably, since then
\begin{equation}
v_{\hat{\theta}}= \frac{k_{2}}{r} \; ,
\end{equation} 
which implies $f_{\hat \theta}=0$. Conversely, if we assume zero
torque, then (assuming $k_1\neq 0$) the vorticity is zero.

But note that the simple relationship \emph{zero torque} $\iff$
\emph{zero vorticity} requires the assumption of nonzero radial flow.
With zero radial flow the torque is always zero for a time independent
flow, regardless of whether or not the flow is vorticity free.

%--------------------------------------------------------------
\paragraph{Zero radial force:}
%--------------------------------------------------------------
Assuming zero radial force, $f_{\hat r}=0$, and assuming $k_1\neq 0$,
one finds
\begin{equation} \label{f_r_5}
{1\over r v_{\hat r}} \;
\left[  {1\over 2} \partial_r [v_{\hat r}(r)^2]
- \frac{v_{\hat \theta}(r)^2}{r} \right]
+ c^2(r) \; {\partial}_r \left( {1\over r v_{\hat r}}\right)  =0 .
\end{equation}
Thus the radial and angular parts of the background velocity are now 
dependent on each other. Once you have chosen, \emph{e.g.},
$v_{\hat{\theta}}(r)$ and $c(r)$, a differential equation constrains
$v_{\hat{r}}$:
\begin{equation} \label{conditon_zero_radial_force}
 v_{\hat{r}} \; \partial_r v_{\hat{r}} \left[1-
   \frac{c^2}{v_{r}^2}\right] 
=  \frac{c^2 + v_{\hat{\theta}}^2}{r} \; .
\end{equation} 

%--------------------------------------------------------------
\paragraph{Zero external force:}
%--------------------------------------------------------------
If we assume zero external force, $\vec f=0$, both the radial and
angular external forces are zero.  Then assuming $k_1\neq0$
\begin{equation}
v_{\hat{r}} \; \partial_r v_{\hat{r}} \left[1-  \frac{c^2}{v_{r}^2}\right] = 
\frac{c^2}{r}+\frac{k_{2}^2}{r^3}\; .
\end{equation} 
Since $c$ depends on $\rho$ via the barotropic equation, and $\rho$
depends on $v_r$ via the continuity equation, this is actually a
rather complicated nonlinear ODE for $v_r(r)$. (For zero radial flow
this reduces to the tautology $0=0$ and we must adopt the analysis of
the previous subsection.)

%--------------------------------------------------------------
%--------------------------------------------------------------

In general the ergosurface is defined by the location where the flow
goes supersonic
\begin{equation}
v_{\hat r}^2 + v_{\hat\theta}^2 = c^2
\end{equation}
while the horizon is defined by the location where
\begin{equation}
v_{\hat r}^2 =  c^2
\end{equation}
Note that horizons can form in three rather different ways:
\begin{itemize}
\item $v_{\hat r} = -c \neq 0$ --- a black hole horizon.
\item $v_{\hat r} = +c\neq 0$ --- a white hole horizon.
\item $v_{\hat r} = \pm c = 0$ --- a bifurcate horizon.
\end{itemize}
In analogue models it is most usual to keep $c\neq0$ and use fluid
flow to generate the horizon, this is the case for instance in the
Painl\'eve--Gullstrand version of the Schwarzschild
geometry~\cite{Visser2,ABH}. The alternate possibility of letting both
$v_{\hat r}\to0$ and $c\to0$ to obtain a horizon is not the most
obvious construction from the point of view of acoustic geometries,
but cannot \emph{a priori} be excluded on either mathematical or
physical grounds. Indeed, it is this less obvious manner of
implementing the acoustic geometry that most closely parallels the 
analysis of Schwarzschild black holes in curvature coordinates or 
isotropic coordinates, and 
we shall soon see that this route is preferred
when investigating the Kerr spacetime.

%--------------------------------------------------------------
\section{The role of dimension}
%--------------------------------------------------------------

The role of spacetime dimension in these acoustic geometries is
sometimes a bit surprising and potentially confusing. This is
important because there is a real physical distinction between truly
two-dimensional systems and effectively two-dimensional systems in the
form of three-dimensional systems with cylindrical symmetry.  We
emphasise that in cartesian coordinates the wave equation
\begin{equation} 
\label{wavef_eq}
\frac{\partial}{\partial {x^{\mu}}} 
\left( f^{\mu \nu} \frac{\partial}{\partial {x^{\nu}}}\; \psi \right)=0,
\end{equation}
where
\begin{equation}
f^{\mu \nu}=\left[
\begin{array}{ccc}
-{\rho}/{c^2}            &|& -{\rho}\;{v^j} /{c^2}\\
%&|& \\
\hline
%&|& \\
-{\rho}\; {v^i}/{c^2}  &|& 
\rho \;\{\delta^{ij} - v^i v^j /{c^2}\}
\end{array}
\right],
\end{equation}
holds \emph{independent} of the dimensionality of spacetime. It
depends only on the Euler equation, the continuity equation, a
barotropic equation of state, and the assumption of irrotational
flow~\cite{Visser1}.

Introducing the acoustic metric $g^{\mu \nu}$, defined by
\begin{equation}
f^{\mu \nu}=\sqrt{-g} \, g^{\mu \nu}; 
\qquad
g=\frac{1}{\det(g^{\mu \nu})}
\end{equation} 
the wave equation (\ref{wavef_eq}) corresponds to the massless
Klein--Gordon equation [d'Alembertian wave equation] in a curved
space-time with contravariant metric tensor:
\begin{equation}
g^{\mu \nu}=\left({\rho\over c}\right)^{-2/(d-1)}
\left[
\begin{array}{c|c}
-1/c^2            & -\vec{v}^{\,T}/c^2 \\
\hline
%--- & & ---------- \\
\vphantom{\Big|}
- \vec{v}/c^2  & 
\mathbf{I}_{d\times d} - \vec{v}\otimes \vec{v}^{\,T}/c^2
\end{array}
\right],
\end{equation}
where $d$ is the dimension of \emph{space} (not spacetime).

The covariant acoustic metric is
\begin{equation}
g_{\mu \nu}=\left( \frac{\rho}{c} \right)^{2/(d-1)}
\left[
\begin{array}{ccc}
-\left( c^2-v^2 \right)       &|& -\vec{v}^{\,T} \\
\hline
%--- & & ---------- \\
-\vec{v}  &|& \mathbf{I}_{d\times d}
\end{array}
\right]. 
\end{equation}

%--------------------------------------------------------------
\subsection{$d=3$}
%--------------------------------------------------------------

The acoustic line-element for three space and one time dimension reads
\begin{equation} 
g_{\mu \nu}=\left( \frac{\rho}{c} \right)
\left[
\begin{array}{ccc}
-\left( c^2-v^2 \right)       &|& -\vec{v}^{\,T} \\
\hline
%--- & & ---------- \\
-\vec{v}  &|& \mathbf{I}_{3\times 3}
\end{array}
\right]. 
\end{equation}
This is the primary case of interest in this article.

%--------------------------------------------------------------
\subsection{$d=2$}
%--------------------------------------------------------------

The acoustic line-element for two space and one time dimension reads
\begin{equation} 
g_{\mu \nu}=\left( \frac{\rho}{c} \right)^2
\left[
\begin{array}{ccc}
-\left( c^2-v^2 \right)       &|&- \vec{v}^{\,T} \\
\hline
%--- & & ---------- \\
-\vec{v}  &|& \mathbf{I}_{2 \times 2}
\end{array}
\right]. 
\end{equation}
This situation would be appropriate when dealing with surface waves or
excitations confined to a particular substrate, for example at the
surface between two superfluids. An important physical point is that,
due to the fact that one can always find a conformal transformation to
make the two-dimensional
spatial slice flat~\cite{Wald}, essentially
all possible 2+1 metrics can in principle be reproduced by acoustic
metrics~\cite{Droplet}.
 (The only real restriction is the quite
physical constraint that there be no closed timelike curves in the
spacetime.)

%--------------------------------------------------------------
\subsection{$d=1$}
%--------------------------------------------------------------

The naive form of the acoustic metric in (1+1) dimensions is
ill-defined, because the conformal factor is raised to a formally
infinite power --- this is a side effect of the well-known conformal
invariance of the Laplacian in 2 dimensions. The wave equation in
terms of $f^{\mu\nu}$ continues to make good sense --- it is only the
step from $f^{\mu\nu}$ to the effective metric that breaks down.

Note that this issue only presents a difficulty for physical systems
that are intrinsically one-dimensional. A three-dimensional system
with plane symmetry, or a two-dimensional system with line symmetry,
provides a perfectly well behaved model for (1+1) dimensions, as in
the cases $d=3$ and $d=2$ above.

%--------------------------------------------------------------
\section{The Kerr equator}
%--------------------------------------------------------------

To compare the vortex acoustic geometry to the physical Kerr geometry
of a rotating black hole~\cite{Kerr}, consider the equatorial slice
$\theta=\pi/2$ in Boyer--Lindquist coordinates~\cite{BL}:
\begin{equation}
\fl
(\d s^2)_{(2+1)} = -\d t^2 + {2m\over r} (\d t - a \;\d\phi)^2 
+{\d r^2\over 1-2m/r+a^2/r^2} +  (r^2+a^2)\;\d\phi^2.
\end{equation}
We would like to put this into the form of an ``acoustic metric''
\begin{equation}
g_{\mu\nu} = \left({\rho\over c}\right)\;
\left[\begin{array}{c|c}{-\{c^2-h^{mn}\;v_n\;v_n}\} & -v_j\\
\hline
-v_i & h_{ij} 
\end{array}\right].
\end{equation}
If we look at the 2-d $r$-$\phi$ plane, the metric is
\begin{equation}
(\d s^2)_{(2)} = 
{\d r^2\over 1-2m/r+a^2/r^2} +  \left(r^2+a^2+ {2ma^2\over r}\right)\d\phi^2.
\end{equation}
Now it is well-known that any 2-d geometry is locally conformally
flat~\cite{Wald}, though this fact is certainly not manifest in these
particular coordinates.  Introduce a new radial coordinate $\tilde r$
such that:
\begin{equation}
{\d r^2\over 1-2m/r+a^2/r^2} +  \left(r^2+a^2+ {2ma^2\over r}\right)\d\phi^2
=
\Omega(\tilde r)^2 \; [\d\tilde r^2 +\tilde r^2\;\d\phi^2].
\end{equation}
This implies
\begin{equation}
 \left(r^2+a^2+ {2ma^2\over r}\right) = \Omega(\tilde r)^2 \; \tilde r^2,
\end{equation}
and
\begin{equation}
{\d r^2\over 1-2m/r+a^2/r^2} = \Omega(\tilde r)^2\; \d\tilde r^2,
\end{equation}
leading to the differential equation
\begin{equation}
{1\over \tilde r(r)}\; {\d \tilde r(r)\over\d r} =
 {1\over \sqrt{1-2m/r+a^2/r^2} \sqrt{r^2+a^2+ {2ma^2/ r}}},
\end{equation}
which is formally solvable as
\begin{equation}
\tilde r(r) = \exp\left\{
\lint  {\d r\over \sqrt{1-2m/r+a^2/r^2} \sqrt{r^2+a^2+ {2ma^2/r}}}
\right\}.
\end{equation}
The normalization is most easily fixed by considering the $m=0=a$
case, in which case $\tilde r=r$, and then using this to write the
general case as
\begin{equation}
\fl
\tilde r(r) = r \exp\left[ -
\int_r^\infty  \left\{
{1\over \sqrt{1-2m/\bar r+a^2/\bar r^2} \sqrt{\bar r^2+a^2+ {2ma^2/\bar r}}} 
-{1\over \bar r} \right\} \d\bar r
\right],
\end{equation}
where $\bar r$ is simply a dummy variable of integration.  If $a=0$
this integral can be performed in terms of elementary functions
\begin{equation}
\int  {\d r\over r \sqrt{1-2m/r}} = \ln \left( \sqrt{r^2 - 2mr} + r - m \right)
\qquad\qquad [r>2m],
\end{equation}
so that
\begin{equation}
\tilde{r} (r)= \half \left( \sqrt{r^2 - 2mr} + r - m \right),
\end{equation}
though for $m\neq 0$ and general $a$ no simple analytic form holds.
Similarly, for $m=0$ and $a\neq 0$ it is easy to show that
\begin{equation}
\tilde r(r) = \sqrt{r^2+a^2}, 
\end{equation}
though for $a\neq 0$ and general $m$ no simple analytic form holds.  

Nevertheless, since we have an exact [if formal] expression for
$\tilde r(r)$ we can formally invert it to assert the existence of a
function $r(\tilde r)$. It is most useful to write $\tilde r=r
\;F(r)$, with $\lim_{r\to\infty}F(r)=1$, and to write the inverse
function as $r = \tilde r \; H(\tilde r)$ with the corresponding limit
$\lim_{\tilde r\to\infty}H(\tilde r)=1$. Even if we cannot write
explicit closed form expressions for $F(r)$ and $H(\tilde r)$ there is
no difficulty in calculating them numerically, or in developing series
expansions for these quantities, or even in developing graphical
representations.

We now evaluate the conformal factor as
\begin{equation}
\Omega(\tilde r)^2 = 
{r^2+a^2+ {2ma^2/ r} \over \tilde r^2} = 
H(\tilde r)^{2} \; \left(1+ {a^2\over r^2} + {2ma^2\over r^3}\right),
\end{equation}
with $r$ considered as a function of $\tilde r$, which now yields
\begin{equation}
(\d s^2)_{(2+1)} = -\d t^2 + {2m\over r} (\d t^2  - 2a \;\d\phi\;\d t) +
\Omega(\tilde r)^2 \; [\d\tilde r^2 +\tilde r^2\;\d\phi^2].
\end{equation}
Equivalently
\begin{eqnarray}
(\d s^2)_{(2+1)} &=& \Omega(\tilde r)^2 
\Bigg\{ -\Omega(\tilde r)^{-2} \left[1- {2m\over r}\right]\d t^2  
-\Omega(\tilde r)^{-2}  {4am\over r} \;\d\phi\;\d t 
\nonumber\\
&&
\qquad\qquad
+ [\d\tilde r^2 +\tilde r^2\;\d\phi^2] \Bigg\}.
\end{eqnarray}
This now lets us pick off the coefficients of the equivalent acoustic
metric.  For the overall conformal factor
\begin{equation}
{\rho\over c} = \Omega(\tilde r)^2 = 
H^2(\tilde r) \; \left(1+ {a^2\over r^2} + {2ma^2\over r^3}\right).
\end{equation}
For the azimuthal ``flow''
\begin{equation}
v_\phi =  \Omega(\tilde r)^{-2} \; {2am\over r} 
=  {2am\over r} \; H^{-2}(\tilde r) \; 
\left(1+ {a^2\over r^2} + {2ma^2\over r^3}\right)^{-1}.
\end{equation}
In terms of orthonormal components
\begin{equation}
v_{\hat\phi} = {v_\phi\over\tilde r}  =  
  {2am\over r^2} \; H^{-1}(\tilde r) \; 
\left(1+ {a^2\over r^2} + {2ma^2\over r^3}\right)^{-1}.
\end{equation}
This is, as expected, a vortex geometry. 
Finally for the ``coordinate speed of light'', corresponding to the
speed of sound in the analogue geometry
\begin{eqnarray}
c^2 &=& \Omega(\tilde r)^{-2} \;\left[1- {2m\over r}\right] + 
\Omega(\tilde r)^{-4} \; {4a^2m^2\over\tilde r^2 \; r^2}, 
\\
&=&  \Omega(\tilde r)^{-4} \left\{ 
\Omega(\tilde r)^{2}\;\left[
1- {2m\over r}\right] +  \; {4a^2m^2\over\tilde r^2 \; r^2} 
\right\}.
\end{eqnarray}
The speed of sound can be rearranged a little
\begin{equation}
c^2 
=  \Omega(\tilde r)^{-4} \;H^2(\tilde r) \; \left\{ 
\left[1+ {a^2\over r^2} + {2ma^2\over r^3}\right]
\;\left[1- {2m\over r}\right] +  \; {4a^2m^2\over r^4} 
\right\}.
\end{equation}
This can now be further simplified to obtain
\begin{equation}
c^2 
=  \Omega(\tilde r)^{-4} \;H^2(\tilde r) \; 
\left\{ 1- {2m\over r} + {a^2\over r^2}
\right\},
\end{equation}
and finally leads to
\begin{equation}
\rho  = H(\tilde r) \; \sqrt{ 1- {2m\over r} + {a^2\over r^2} },
\end{equation}
with $r$ implicitly a function of $\tilde r$. Note that the velociy field of the ``equivalent Kerr vortex'' is not irrotational. Application of Stokes' theorem quickly yields
\begin{equation}
\fl
\omega
= {1\over 2\pi \tilde r} {\partial\over\partial\tilde r}\left(2\pi \tilde r \; v_{\hat\phi} \right)
= {1\over \tilde r} {\partial\over\partial\tilde r}\left( v_{\phi} \right)
= {1\over \tilde r} {\partial\over\partial\tilde r}\left[ {2am \tilde r^2\over r^3} \; 
\left(1+ {a^2\over r^2} + {2ma^2\over r^3}\right)^{-1}\right].
\end{equation}
which we can somewhat simplify to
\begin{eqnarray}
\omega
&= {4am \over r^3} \; 
\left(1+ {a^2\over r^2} + {2ma^2\over r^3}\right)^{-1} 
\\
&
+ {\Omega\;\tilde r}\; \sqrt{1-2m/r+a^2/r^2} \;
 {\partial\over\partial r}\left[ {2am \over r^3} \; 
\left(1+ {a^2\over r^2} + {2ma^2\over r^3}\right)^{-1}\right].
\nonumber
\end{eqnarray}
Note that (as should be expected), the vorticity is proportional to an 
overall factor of $a$, and so is proportional to the angular momentum 
of the Kerr geometry we are ultimately interested in.
%%% ***

While $\tilde r$ is the radial coordinate in which the space part of
the acoustic geometry is conformally flat, so that $\tilde r$ is the
``physical'' radial coordinate that corresponds to distances measured
in the laboratory where the vortex has been set up, this particular
radial coordinate is also mathematically rather difficult to work
with.  For some purposes it is more useful to present the coefficients
of the acoustic metric as functions of $r$, using the relationship
$F(r) = 1/H(\tilde r)$. Then we have:
\begin{equation}
{\rho(r)\over c(r)} = \Omega^2(r) = 
F^{-2}(r) \; \left(1+ {a^2\over r^2} + {2ma^2\over r^3}\right).
\end{equation}
\begin{equation}
v_{\hat\phi}(r)
=   {2am\over r^2} \; F(r) \; 
\left(1+ {a^2\over r^2} + {2ma^2\over r^3}\right)^{-1}.
\end{equation}
\begin{equation}
c(r)
=  \sqrt{ 1- {2m\over r} + {a^2\over r^2} } \; F(r) \;   
\left(1+ {a^2\over r^2} + {2ma^2\over r^3}\right)^{-1}.
\end{equation}
Finally this yields
\begin{equation}
\rho(r) = F^{-1}(r) \; \sqrt{ 1- {2m\over r} + {a^2\over r^2} },
\end{equation}
and the explicit if slightly complicated result that
\begin{equation}
\fl
F(r) = \exp\left[ -
\int_r^\infty  \left\{
{1\over \sqrt{1-2m/\bar r+a^2/\bar r^2} \sqrt{\bar r^2+a^2+ {2ma^2/\bar r}}} 
-{1\over \bar r} \right\} \d\bar r
\right].
\end{equation}
One of the advantages of writing things this way, as functions of $r$,
is that it is now simple to find the locations of the horizon and
ergosphere. 

The ergo-surface is defined by $c^2-v^2=0$, equivalent to the
vanishing of the $g_{tt}$ component of the metric. This occurs at
\begin{equation}
r_E=2m.
\end{equation}
The horizon is determined by the vanishing of $c$, (recall that $v_{\hat
  r}\equiv0$), this requires solving a simple quadratic with the result
\begin{equation}
r_H =  m+ \sqrt{m^2-a^2} < r_E.
\end{equation}
These results agree, as of course they must, with standard known
results for the Kerr metric. It is easy to check that $F(r_H)$ is
finite (the function inside the exponential is integrable as long as
the Kerr geometry is non-extremal), and that $F(r_E)$ is finite. This
then provides some simple consistency checks on the geometry:
\begin{eqnarray}
\rho(r_H)&=&0 \, =  \, c(r_H);
\\
\Omega(r_H)&=& \hbox{finite};
\\
F(r_H) &=& G(r_H)^{-1} =  \hbox{finite};
\\
\Omega(r_H) \; v_{\hat\phi}(r_H) &=& {a\over r_H};
\\
\omega(r_H) &=& {a\over 2 m r_H};
\\
c(r_E)&=& |v_{\hat\phi}(r_E)|.
\end{eqnarray}
In particular if we compare the vorticity at the horizon, 
$\omega(r_H) = {a/( 2 m r_H)}$, with the surface gravity 
\begin{equation}
\kappa_H = {2m\over r_H^2} -{2a^2\over r_H^3},
\end{equation}
we find a large range of parameters for which the peak frequency 
of the Hawking spectrum is high compared to the frequency scale 
set by the vorticity --- this acts to suppress any complications 
specifically coming from the vorticity. (Of course in the extremal 
limit, where $r_H\to 0$, the vorticity at the horizon becomes infinite.
In the extremal limit one would need to make a more careful analysis 
of the effect of nonzero vorticity on the Hawking radiation.)
%%% ***

From the point of view of the acoustic analogue the region inside the
horizon is unphysical.  In the unphysical region the density of the 
fluid is zero and the concept of sound meaningless.  In the physical 
region outside the horizon the flow has zero radial velocity, and zero torque, but
is \emph{not} irrotational.  Since by fitting the equatorial slice of
Kerr to a generic acoustic geometry we have fixed $\rho(r)$ and
$c^2(r)$ as functions of $r$ [and so also as functions of $\tilde r$]
it follows that $p(r)$ is no longer free, but is instead determined by
the geometry. From there, we see that the EOS $p(\rho)$ is determined,
as is the external force $f_{\hat r}(r)$.  The net result is that we
can [in principle] simulate the Kerr equator exactly, but at the cost
of a very specific \emph{fine-tuning} of both the equation of state
$p(\rho)$ and the external force $f_{\hat r}(r)$.

\begin{figure}[htbp]
\vbox{\hfil\scalebox{0.50}{{\includegraphics{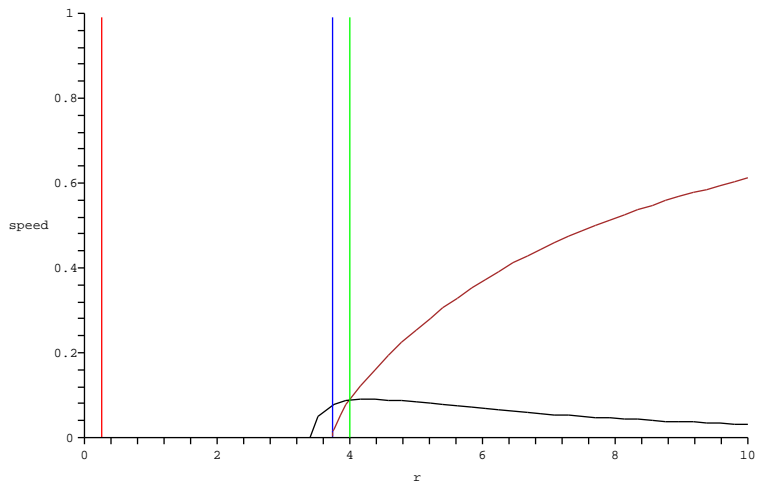}}}\hfil}
\bigskip
\caption{
%------------------------------
{\sl Illustrative graph of the speed of sound and rotational velocity
  of the vortex as a function of $r$. ($m=2$, $a=1$, 30'th-order
  Taylor series.)}
%------------------------------
}
\label{F:cv}
\end{figure}

In figure \ref{F:cv} we present an illustrative graph of the speed of
sound and velocity of the fluid.  Key points to notice are that the
speed of sound (brown curve) and velocity of flow (black curve)
intersect on the ergosurface (green line), that the speed of sound
goes to zero at the horizon (blue line), and that the velocity of
fluid flow is finite on the horizon. Finally the inner horizon is
represented by the red vertical line. While the region inside the
outer horizon is not physically meaningful, one can nevertheless
mathematically extend some (not all) of the features of the flow
inside the horizon.  This is another example of the fact that analytic
continuation across horizons, which is a standard tool in general
relativity, can sometimes fail for physical reasons when one considers
Lorentzian geometries based on physical models that differ from
standard general relativity. This point is more fully discussed in
\cite{causal,JV,JK}.

\begin{figure}[htbp]
\vbox{\hfil\scalebox{0.50}{{\includegraphics{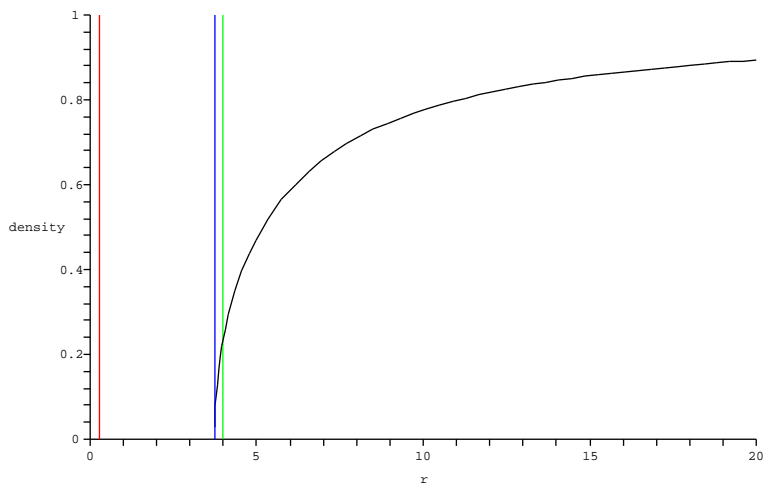}}}\hfil}
\bigskip
\caption{
%------------------------------
 {\sl Illustrative graph of the fluid density as a function of
   $r$. ($m=2$, $a=1$, 30'th-order Taylor
   series.)}
%------------------------------
}
\label{F:density}
\end{figure}

\begin{figure}[htbp]
\vbox{\hfil\scalebox{0.50}{{\includegraphics{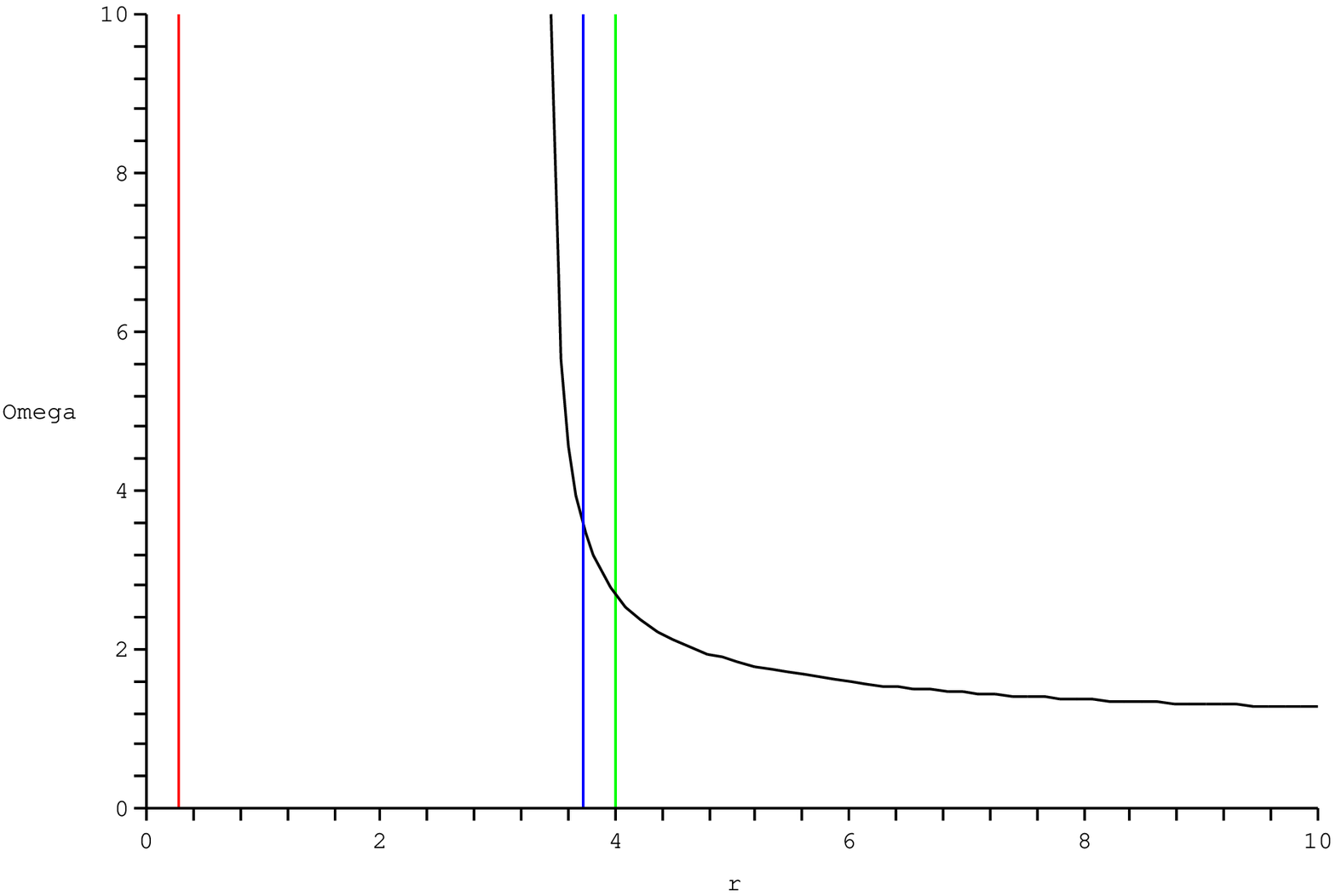}}}\hfil}
\bigskip
\caption{
%------------------------------
 {\sl Illustrative graph of the conformal factor $\Omega$ as a
   function of $r$. ($m=2$, $a=1$, 30'th-order Taylor series.)}
%------------------------------
}
\label{F:Omega}
\end{figure}

In figure \ref{F:density} we present an illustrative graph of the
fluid density. Note that the density asymptotes to a constant at large
radius, and approaches zero at the horizon. Finally, in figure
\ref{F:Omega} we present an illustrative graph of the conformal factor
$\Omega$, which remains finite at the horizon and ergosurface, and
asymptotes to unity at large distances from the vortex core. Figures
\ref{F:cv}--\ref{F:Omega} all correspond to $m=2$ and $a=1$, and have
been calculated using a 30'th-order Taylor series
expansion.~\footnote{A suitable {\sf Maple} worksheet is available from
  the authors.}

%=====================================================================
%=====================================================================

%=====================================================================
%=====================================================================

%=====================================================================
%=====================================================================

%-------------------------------------------------------------------
\section{Discussion}
%-------------------------------------------------------------------

We have shown that the Kerr equator can [in principle] be exactly
simulated by an acoustic analogue based on a vortex flow with a very
specific equation of state and subjected to a very specific external
force. Furthermore we have as a result of the analysis also seen that
such an analogue would have to be very specifically and deliberately
engineered. Thus the results of this investigation are to some extent
mixed, and are more useful for theoretical investigations, and for the
gaining of insight into the nature of the Kerr geometry, than they are
for actual laboratory construction of a vortex simulating the Kerr
equator.

One of the technical surprises of the analysis was that the
Doran~\cite{Doran,River} coordinates (the natural generalization of the
Painl\'eve--Gullstrand coordinates that worked so well for the
Schwarzschild geometry) did not lead to a useful acoustic metric even
on the equator of the Kerr spacetime.  Ultimately, this can be traced
back to the fact that in Doran coordinates the space part of the Kerr
metric is non-diagonal, even on the equator, and that no simple
coordinate change can remove the off-diagonal elements. In
Boyer--Lindquist coordinates however the space part of the metric is
at least diagonal, and the $\tilde r$ coordinate introduced above
makes the spatial part of the equatorial geometry conformally
flat. This $\tilde r$ coordinate is thus closely related to the radial
coordinate in the isotropic version of the Schwarzschild solution and
$\tilde r$ in fact reduces to that isotropic radial coordinate as
$a\to0$.

However, in a more general physical setting the
Doran~\cite{Doran,River} coordinates might still be useful for
extending the acoustic analogue away from the equator, and into the
bulk of the domain of outer communication.  To do this we would have
to extend and modify the notion of acoustic geometry. The fact that
the spatial slices of the Kerr geometry are \emph{never} conformally
flat~\cite{Price,Valiente-Kroon1,Valiente-Kroon2} forces any attempt
at extending the acoustic analogy to consider a more general class of
acoustic spacetimes. A more general physical context that may prove
suitable in this regard is anisotropic fluid media, and we are currently 
investigating this possibility. The simplest anisotropic fluid media are 
the classical liquid crystals~\cite{liquid1,liquid2}, which in the present 
context suffer from the defect that they possess considerable viscosity and exist chiefly at room temperature --- this makes then unsuitable for the construction of analogue horizons, and particularly unsuitable for the investigation of quantum aspects of analogue horizons.  Much more promising in this regard are the quantum liquid crystals. These are anisotropic superfluids such as, in particular, 3He--A~\cite{Droplet,Volovik-report}.  The anisotropic superfluids are generally characterized by the presence of a non-scalar order parameter; they behave as superfluid liquid crystals, with zero friction at $T=0$. The premier example of an anisotropic superfluid is fermionic  3He--A, where the effective gravity for fermions is described by vierbeins, though 
several types of Bose and Fermi condensates of cold atoms also exhibit similar behaviour~\cite{Droplet,Volovik-report}.

%%% ***

In the system we have presented above we have found that two of the
main features of the Kerr geometry appear: the horizon (outer horizon)
and ergo-region.  By slight modification of standard arguments, these geometric features are expected to lead
to the quantum phenomena of super-radiance and Hawking
radiation~\cite{Unruh1,Unruh2,Visser1,Visser2,ABH}. The super-radiance
corresponds to the reflection and amplification of a wave packet at
the ergo surface~\cite{Savage}. (See also~\cite{Basak1,Basak2}.) The
model, because it contains a horizon, also satisfies the basic
requirements for the existence of Hawking radiation, which would now
be a thermal bath of phonons emitted from the
horizon~\cite{Unruh1}.~\footnote{Furthermore, there is a strong
  feeling that the cosmological constant problem in elementary
  particle physics might be related to a mis-identification of the
  fundamental degrees of freedom~\cite{Droplet,Volovik-report}.} These
are two of the most fundamental physics reasons for being interested
in analogue models~\cite{ABH}.
A subtlety in the argument that leads to Hawking radiation arises from the way that the choice of coordinates \emph{seems} to influence the choice of quantum vacuum state. In the presence of a horizon, the choice of quantum vacuum is no longer unique, and standard \emph{choices} for the quantum vacuum are the Unruh, Hartle--Hawking, and Boulware states. It is the Unruh vacuum that corresponds (for either static or stationary black holes) to Hawking radiation into an otherwise empty spacetime, while the Hartle--Hawking vacuum (for a static black hole) corresponds to a black hole in thermal equilibrium with its environment (at the Hawking temperature). For a stationary [non-static] black hole the Hartle--Hawking vacuum state does not strictly speaking exist~\cite{Wald-LR}, but there are quasi-Hartle--Hawking quantum states that possess most of the relevant features~\cite{Iyer-Kumar}. In general relativity, because physics is coordinate independent, the choice of vacuum state is manifestly independent of the choice of coordinate system. In the analogue spacetimes considered in this article, because the preferred choice of coordinate system is intimately tied to the flowing medium, the situation is perhaps less clear.  The coordinate system we have adopted is invariant under combined time reversal and parity, which might tempt one to feel that a pseudo-Hartle--Hawking vacuum is the most natural one. On the other hand, if we think of building up the vortex from a fluid that is initially at rest, then there is clearly a preferred time direction, and the ``white hole'' component of the maximally extended horizon would exist only as a mathematical artifact.  In this physical situation the Unruh vacuum is the most natural one. We feel that there is an issue worth investigating here,  to which we hope to return in some future article.

%%% ***

Finally, since astrophysically all black holes are expected to exhibit some
degree of rotation, it is clear that an understanding of the influence
of rotation on analogue models is important if one wishes to connect
the analogue gravity programme back to astrophysical
observations.~\footnote{An interesting attempt in the opposite
  direction~\cite{das1,das2} is the recent work that analyzes
  accretion flow onto a black hole in terms of a dumb hole
  superimposed upon a general relativity black hole.}  In conclusion,
there are a number of basic physics reasons for being interested in
acoustic analogues of the Kerr geometry, and a number of interesting
directions in which the present analysis might be extended.

%------------------------------------------------------------------
\section*{Acknowledgements}
%------------------------------------------------------------------

This research was supported by the Marsden Fund administered by the
Royal Society of New Zealand.

%\appendix
%\clearpage
%------------------------------------------------------------------
\section*{Appendix: \\
Isotropic version of the Schwarzschild geometry cast in acoustic form}
%------------------------------------------------------------------

The isotropic version of the Schwarzschild geometry can be cast
\emph{exactly} into acoustic form, without any extraneous conformal
factors. Since this is not the standard way of viewing the acoustic
analogue of the Schwarzschild geometry~\cite{Visser2}, and since the algebra is
simple enough to be done in closed form, it is worthwhile taking a
small detour.

In isotropic coordinates the Schwarzschild geometry reads
\begin{equation}
\d s^2 = - {(1-{m\over2r})^2\over(1+{m\over2r})^2}\;\d t^2 
+ \left(1+{m\over2r}\right)^4 
\left\{ \d r^2 + r^2(\d\theta^2 +\sin^2\theta\;\d\phi^2) \right\},
\end{equation}
and so in these coordinates the acoustic analogue corresponds to
\begin{eqnarray}
v &=& 0;
\\
\rho &=& \rho_\infty\;\left(1 - {m^2\over 4r^2} \right);
\\
c^2 &=&   c^2_\infty\; {(1-{m\over2r})^2\over(1+{m\over2r})^6}.
\end{eqnarray}
The external force required to hold this configuration in place
against the pressure gradient is
\begin{equation}
f_{\hat r} = {\d p\over\d r} = c^2 \; {\d\rho\over\d r} =  
\rho_\infty \; c^2_\infty \; 
{(1-{m\over2r})^2\over(1+{m\over2r})^6} \;  {m^2\over 2 r^3}.
\end{equation}
The pressure itself (normalizing $p_\infty\to0$) is then
\begin{equation}
p(r) =  -  \rho_\infty \; c^2_\infty \;  m^2 \; 
\left({ 120 r^3 - 20 r^2 m+10m^2r+m^3\over 480\;r^5\;(1+{m\over2r})^5} \right),
\end{equation}
and by eliminating $r$ in favour of $\rho$ using
\begin{equation}
r = {m\over2\sqrt{1-\rho/\rho_\infty}}
\end{equation}
we can deduce the equation of state
\begin{equation}
\fl
p(\rho) = - (\rho_\infty-\rho)\;   c^2_\infty \;
\left({ 
20-5\rho/\rho_\infty-4\sqrt{1-\rho/\rho_\infty}
-\rho/\rho_\infty\sqrt{1-\rho/\rho_\infty}
\over 
15\; (1+\sqrt{1-\rho/\rho_\infty})^5}
\right).
\end{equation}
At the horizon, which occurs at $r_H=m/2$ in these coordinates, both
$\rho=0$ and $c=0$ in the acoustic analogue, while
\begin{equation}
p_H =  -{ \rho_\infty \; c^2_\infty \over 30}
\end{equation}
is finite, so there is a finite pressure drop between asymptotic
infinity and the horizon. Everything is now simple enough to be fully
explicit, and although we now have an analogue model that reproduces
the \emph{exterior} part of the Schwarzschild geometry \emph{exactly},
we can also clearly see the two forms in which fine tuning arises ---
in specifying the external force, and in the equation of state.
 
More generally, it is a standard result that \emph{any} spherically
symmetric geometry can be put into isotropic form
\begin{equation}
\d s^2 = - \exp[-2\Phi]\;\d t^2 + \exp[-2\Psi]
\left\{ \d r^2 + r^2(\d\theta^2 +\sin^2\theta\;\d\phi^2) \right\}.
\end{equation}
This can be put into acoustic form in a particularly simple manner by
setting
\begin{eqnarray}
v(r) &=& 0;
\\
\rho(r) &=& \rho_\infty\; \exp[-\Phi-\Psi];
\\
c(r) &=&   c_\infty\;  \exp[\Psi-\Phi].
\end{eqnarray}
The pressure can now be formally evaluated as
\begin{equation}
p(r) =   \rho_\infty \; c_\infty^2 \int_r^\infty \exp[\Psi-3\Phi] \; \partial_r (\Phi+\Psi) \; \d r,
\end{equation}
and comparison with $\rho(r)$ can be used to construct a formal equation of state, $\rho(p)$.

%------------------------------------------------
\section*{References}
%------------------------------------------------

%----------------------------------------------------------------------

%-----------------------------------------------------------------------

%-----------------------------------------------------------------------
\end{document}